\documentclass[preprint,authoryear,12pt]{elsarticle}

\usepackage{epsfig}
\usepackage[usenames, dvipsnames]{color}

\usepackage{amssymb}

\usepackage[
a4paper=true,%
breaklinks=true,%
colorlinks=true,%
pdfauthor={Fan et al.},%
pdftitle={Template for manuscripts in Advances in Space Research}%
]{hyperref}

\journal{Advances in Space Research}

\begin{document}

%%%%%%%%%%%%%%%%%%%%%%%%%%%%%%%%%%%%%%%%%%%%%%%%%%%%%%%%%%%%%%%%%%%%%%%%%%%%%
%% Frontmatter
\begin{frontmatter}


% Use the tnoteref command within \title and fnref within \author or \address for footnotes;
% use the corref command within \author for corresponding author footnotes;
% \author{Name\corref{cor1}\fnref{label2}}
% \address{Address\fnref{label3}}

\title{Differential Rotation in Solar Convective Dynamo Simulations}
%%\tnotetext[footnote1]{This template can be used for all publications in Advances in Space Research.}

% Use optional labels to link authors explicitly to addresses:

\author{Yuhong Fan\fnref{footnote1}}
\address{High Altitude Observatory, National Center for Atmospheric Research, 3280 Center Green Dr., Boulder, CO 80301, USA}
%\cortext[cor]{Corresponding author}
\fntext[footnote1]{Corresponding author}
\ead{yfan@ucar.edu}

% Url can be given like this:
% \ead[url]{http://www.elsevier.com/wps/find/authorsview.authors/latex}

\author{Fang Fang}
\address{Laboratory For Atmospheric and Space Physics, University of Colorado
at Boulder, 1234 Innovation Dr., Boulder, CO, 80303, USA}
\ead{fang.fang-1@colorado.edu}

%%\author{More Authors\fnref{footnote4}}
%%\address{Address of the co-authors}
%%\fntext[footnote4]{Additional information about the co-authors}
%%\ead{more@email.addresses}

\begin{abstract}
We carry out a magneto-hydrodynamic (MHD) simulation of convective dynamo in
the rotating solar convective envelope driven by the solar radiative diffusive
heat flux.  The simulation is similar to that reported in
\citet{Fan:Fang:2014} but with further reduced viscosity and magnetic diffusion.
The resulting convective dynamo produces a large scale mean field that 
exhibits similar irregular cyclic behavior and polarity reversals, and
self-consistently maintains a solar-like differential rotation.
The main driver for the solar-like differential rotation (with faster rotating
equator) is a net outward transport of angular momentum away from the rotation
axis by the Reynolds stress, and we found that this transport is enhanced with
reduced viscosity and magnetic diffusion.

\end{abstract}


\end{frontmatter}

\parindent=0.5 cm

%%%%%%%%%%%%%%%%%%%%%%%%%%%%%%%%%%%%%%%%%%%%%%%%%%%%%%%%%%%%%%%%%%%%%%%%%%%%%
%% Main text
\section{Introduction}
Significant advances have been made in recent years in global-scale fully
dynamic three-dimensional convective dynamo simulations of the solar/stellar
convective envelopes to reproduce some of the basic features of the Sun's large-scale
cyclic magnetic field \citep[e.g.][]{Ghizaru:etal:2010,
Racine:etal:2011,Kapyla:etal:2012,Nelson:etal:2013,Fan:Fang:2014,
Augustson:etal:2015,Hotta:etal:2015}.
It has also been found that the presence of the dynamo generated magnetic
fields may be important in the self-consistent maintenance of the
solar-like differential rotation with the equator rotating faster than
the poles \citep[e.g.][]{Fan:Fang:2014,Karak:etal:2015,Mabuchi:etal:2015,
Simitev:etal:2015}.
It has been shown that if the buoyancy driving of convection becomes too
strong compared to the Coriolis force such that the convection is no-longer
rotationally constrained, as measured by the
Rossby number $R_{o} \equiv v/\Omega L $ becoming significantly greater
than 1, where $v$ and $L$ are the characteristic convective speed and
length scale and $\Omega$
is the angular rotation rate, anti-solar differential rotation
(faster rotating poles) develops \citep[e.g.][]{Matt:etal:2011,
Gastine:etal:2013, Guerrero:etal:2013}.
\citet{Fan:Fang:2014} found that the dynamo generated magnetic field can
have the effect of an enhanced viscosity that suppresses convection to
keep it rotationally constrained such that the Reynolds stress generated
by the convection drives a solar-like differential rotation, while in the
corresponding hydrodynamic case in the absense of the magnetic field the
differential rotation becomes anti-solar.
Here we continue to investigate the dynamic reaction of the magnetic
field by further reducing the viscosity and the magnetic diffusivity
in the convective dynamo simulation from that of \citet{Fan:Fang:2014}.
We find that the self-consistent maintenance of the solar-like
differential rotation remains robust, and the resulting large-scale
mean magnetic field also shows a similar irregular cyclic behavior.
With reduced viscosity and magnetic diffusivity, we find the tendency
of an enhanced magnetic energy, and a reduction
of the kinetic energy of convection on the larger scales.
As a result the outward
transport (away from the rotation axis) of angular momentum by the
Reynolds stress is further increased, balanced by an increased inward
transport by the magnetic stress, with the transport by the viscous
stress reduced to a negligible level.

\section{Description of the Numerical Simulations}
The setup of the numerical simulation is the same as that described in
\citet[][hereafter FF]{Fan:Fang:2014} except the specification of the
kinetic viscosity $\nu$ and magnetic diffusivity $\eta$.
Here for the clarity of
this paper we reiterate the key features of the numerical model.
We solve the following anelastic MHD equations
using a finite-difference spherical anelastic MHD code
\citep{Fan:etal:2013}:
\begin{equation}
\nabla \cdot ( \rho_0 {\bf v} ) = 0,
\label{eq:continuity}
\end{equation}
\begin{equation}
\rho_0 \left [ \frac{\partial {\bf v}}{\partial t} + ({\bf v} \cdot \nabla )
{\bf v} \right ]
= 2 \rho_0 {\bf v} \times {\bf \Omega}_0 - \nabla p_1 + \rho_1 {\bf g}
+ \frac{1}{4 \pi} ( \nabla \times {\bf B}) \times {\bf B}
+ \nabla \cdot {\cal D} ,
\label{eq:momentum}
\end{equation}
\begin{eqnarray}
\rho_{0} T_0 \left [ \frac{\partial s_1}{\partial t}
 + ( {\bf v} \cdot \nabla ) (s_0 + s_1 ) \right ]
& = & \nabla \cdot ( K \rho_0 T_0 \nabla s_1 )
\nonumber \\
& & - ( {\cal D} \cdot \nabla )
\cdot {\bf v} + \frac{1}{4 \pi} \eta ( \nabla \times {\bf B} )^2
\nonumber \\
& & - \nabla \cdot {\bf F}_{\rm rad},
\label{eq:entropy1}
\end{eqnarray}
\begin{equation}
\nabla \cdot {\bf B} = 0 ,
\label{eq:divB}
\end{equation}
\begin{equation}
\frac{\partial {\bf B}}{\partial t} = \nabla \times ( {\bf v} \times {\bf B} )
- \nabla \times ( \eta \nabla \times {\bf B} ) ,
\label{eq:induction1}
\end{equation}
\begin{equation}
\frac{\rho_1}{\rho_0} = \frac{p_1}{p_0} - \frac{T_1}{T_0},
\label{eq:eqstate}
\end{equation}
\begin{equation}
\frac{s_1}{c_p} = \frac{T_1}{T_0} - \frac{\gamma -1}{\gamma}
\frac{p_1}{p_0} ,
\label{eq:2ndthermlaw}
\end{equation}
where $s_0 (r)$, $p_0 (r)$, $\rho_ 0 (r)$, $T_0 (r)$, and
${\bf g} = - g_0 (r) {\hat {\bf r}}$ are
the profiles of entropy, pressure, density, temperature, and the gravitational
acceleration of a time-independent, reference state of hydrostatic equilibrium
and nearly adiabatic stratification, $c_p$ is the specific heat capacity at
constant pressure, $\gamma$ is the ratio of specific
heats, ${\bf v}$, ${\bf B}$, $s_1$, $p_1$,
$\rho_1$, and $T_1$ are the dependent velocity field, magnetic field,
entropy, pressure, density, and temperature to be solved that describe the
changes from the reference state. Also in the above equations,
${\bf \Omega}_0$ with $\Omega_0 = 2.7 \times 10^{-6} {\rm rad} \; {\rm s}^{-1}$,
denotes the solid body rotation rate of the Sun which is used as the frame
of reference. ${\cal D}$ denotes the viscous stress tensor:
${\cal D}_{ij} = \rho_0 \nu \left [ S_{ij} - (2/3) ( \nabla \cdot
{\bf v} ) \delta_{ij} \right ]$,
where $\nu$ is the kinematic viscosity, $\delta_{ij}$ is the unit tensor, and
$S_{ij}$ is the strain rate tensor given by the following in spherical
polar coordinates: 
\begin{equation}
S_{rr} = 2 \frac{\partial v_r}{\partial r}
\end{equation}
\begin{equation}
S_{\theta \theta} = \frac{2}{r} \frac{\partial v_{\theta}}{\partial \theta}
+ \frac{2 v_r}{r}
\end{equation}
\begin{equation}
S_{\phi \phi} = \frac{2}{r \sin \theta} \frac{\partial v_{\phi}}{\partial \phi}
+ \frac{2 v_r}{r} + \frac{2 v_{\theta}}{r \sin \theta } \cos \theta
\end{equation}
\begin{equation}
S_{r \theta} = S_{\theta r} = \frac{1}{r} \frac{\partial v_r}{\partial \theta}
+ r \frac{\partial}{\partial r} \left ( \frac{v_{\theta}}{r} \right )
\end{equation}
\begin{equation}
S_{\theta \phi } = S_{\phi \theta} = \frac{1}{r \sin \theta}
\frac{\partial v_{\theta}} {\partial \phi } + \frac{\sin \theta} {r}
\frac{\partial}{\partial \theta} \left ( \frac{v_{\phi}}{\sin \theta} \right )
\end{equation}
\begin{equation}
S_{\phi r} = S_{r \phi } = \frac{1}{r \sin \theta}
\frac{\partial v_r}{\partial \phi} + r \frac{\partial}{\partial r}
\left ( \frac{v_{\phi}}{r} \right ) .
\end{equation}
$K$ denotes the thermal diffusivity, and $\eta$ is the magnetic diffusivity.
In equation (\ref{eq:entropy1}),
\begin{equation}
{\bf F_{\rm rad}} = - \frac{ 16 \sigma_s {T_0}^3}{3 \kappa
\rho_0 } \frac{d T_0}{dr} \, {\hat {\bf r}}
\label{eq:frad}
\end{equation}
is the radiative diffusive heat flux,
where $\sigma_s$ is the Stefan-Boltzmann constant,
$\kappa$ is the Rosseland mean opacity.

The simulation domain is a partial spherical shell where $r$ ranges
from $0.722 R_{\odot}$ (base of the solar convection zone) to $0.971 R_{\odot}$
(about 20 Mm below the solar photosphere), $\theta$ (polar angle) ranges
between $\pm 60^{\circ}$ latitudes, and $\phi$ spans the full azimuthal range
of $0^{\circ}$ to $360^{\circ}$. The domain is resolved by
a grid of $96(r) \times 512 (\theta) \times 768 (\phi)$.
We use J. Christensen-Dalsgaard's (JCD) solar model \citep{JCD:etal:1996}
for the reference profiles of $T_0(r)$, $\rho_0 (r)$, $p_0 (r) $, $g_0 (r)$
and assume $s_0 (r) =0$.
The heating $- \nabla \cdot {\bf F}_{\rm rad}$ in equation (\ref{eq:entropy1})
due to the solar radiative diffusive heat flux given by the JCD model drives
a radial gradient of $s_1$ that drives the convection.
In FF, we set the thermal diffusivity
$K = 3 \times 10^{13} \, {\rm cm}^2 \, {\rm s}^{-1}$, the kinetic viscosity $\nu =
10^{12} \, {\rm cm}^2 \, {\rm s}^{-1}$, and the magnetic diffusivity
$\eta = 10^{12} \, {\rm cm}^2 \, {\rm s}^{-1}$ at the top of the domain, and
they all decrease with depth following a $1/ \sqrt{\rho_0}$ profile.
Here in this simulation, we keep the same thermal diffusivity $K$, but
set $\nu$ and $\eta$ both to zero.  Thus the explicit viscous and resistive
terms in the above anelastic MHD equations become zero, but there are still
numerical diffusions due to the upwinded, slope-limited evaluations of 
the fluxes in the advection terms and the electric field in the induction
equation \citep{Fan:etal:2013}.

We use the same boundary and initial conditions as FF. The boundaries
in $r$ and $\theta$ are non-penetrating and stress free for the velocity,
perfect electric conducting walls for the magnetic field at the bottom
and $\theta$ boundaries, and radial magnetic field at the top boundary.
These ensure strict angular momentum conservation for the domain. We impose
$\partial s_1 / \partial r = 0$ at the bottom boundary, $s_1 = 0$ at the
top boundary, and $\partial s_1 / \partial \theta = 0$ at the $\theta$
boundaries. Thus the only energy flux coming into the domain 
is the solar radiative diffusive heat flux through the bottom,
and the only energy flux coming out of the domain is the conductive heat
flux due to the thermal diffusion at the top.
A latitudinal gradient of entropy $s_1$ is also imposed at the lower
boundary, corresponding to a pole to equator temperature
gradient of about $6.8$ K, to represent the tachocline induced
entropy variation that can break the Taylor-Proudman constraint in the
convection zone \citep{Rempel:2005, Miesch:etal:2006}.
The initial state is an unstable thermal equilibrium with a
super-adiabatic initial $\left < s_1 \right >$ profile (with
``$\left < \right >$'' denoting horizontal averaging), where the
thermal conductive heat flux together with the radiative diffusive 
heat flux carries the (constant) solar luminosity through the domain.
We start with a small seed velocity and magnetic field and let the
magneto-convection develop in the domain.

\section{Results}
\label{sec:results}
\begin{figure}
\begin{center}
\includegraphics*[width=10cm,angle=0]{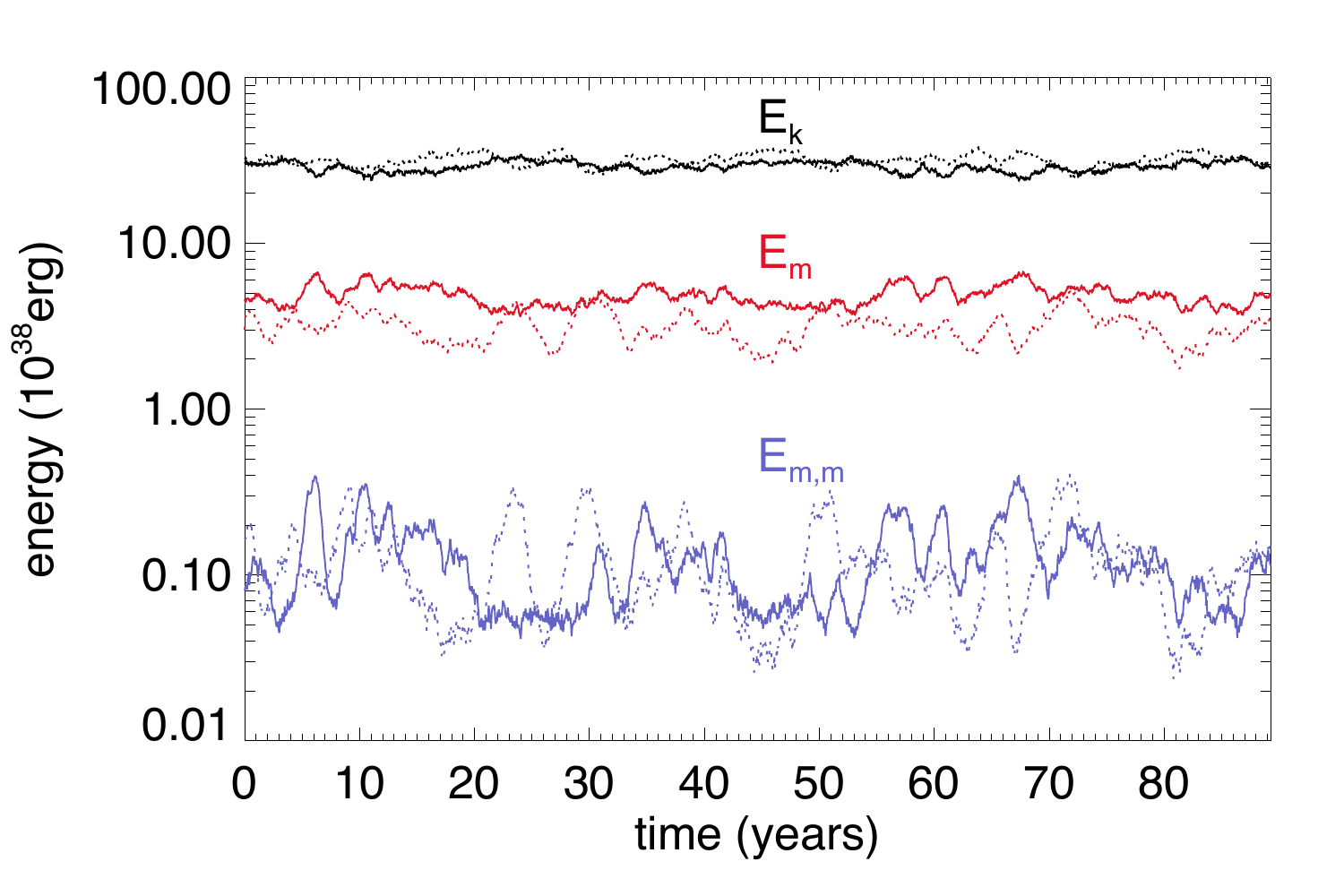}
\end{center}
\caption{The variation of total kinetic energy $E_k$ (black curves), the total magnetic energy $E_m$
(red curves), and the azimuthally averaged mean magnetic energy $E_{\rm m,m}$ (blue curves) over
a sample time span of 89 years of the statistically steady evolution from the present simulation (solid
curves) and, in comparison, those from the dynamo simulation in FF (dotted curves)}
\label{fig:engcompare}
\end{figure}

Figure \ref{fig:engcompare} shows the time sequences of the total kinetic
energy $E_{\rm k}$, the total magnetic energy $E_{\rm m}$, and the azimuthally
averaged mean magnetic energy $E_{\rm m,m}$ over a sample time span of about
89 years after the convective dynamo has reached statistically steady
evolution (solid curves), in comparison to those
obtained in FF (dashed curves).
With reduced magnetic diffusivity and viscosity in the current convective
dynamo simulation, we find that the total
magnetic energy $E_{\rm m}$ is on average enhanced to about 1.51 times
that of FF, while the total kinetic energy $E_{\rm k}$ is on average
suppressed to about 0.91 of that in FF, and the mean (azimuthally averaged)
magnetic energy $E_{\rm m,m}$ amplitude is very close to that of FF, although
on average there is a statistically significant increase (at about the
3 sigma level) compared to FF.
%%\textcolor{red}
{The ratio of the poloidal magnetic energy over the toroidal magnetic energy,
$\int ({B_r}^2+{B_{\theta}}^2) dV / \int {B_{\phi}}^2 dV$, is slightly
increased, being $1.160 \pm 0.001$ in the present case compared to
$1.105 \pm 0.001$ in FF.}

\begin{figure}
\centering
\includegraphics[width=10cm,angle=0]{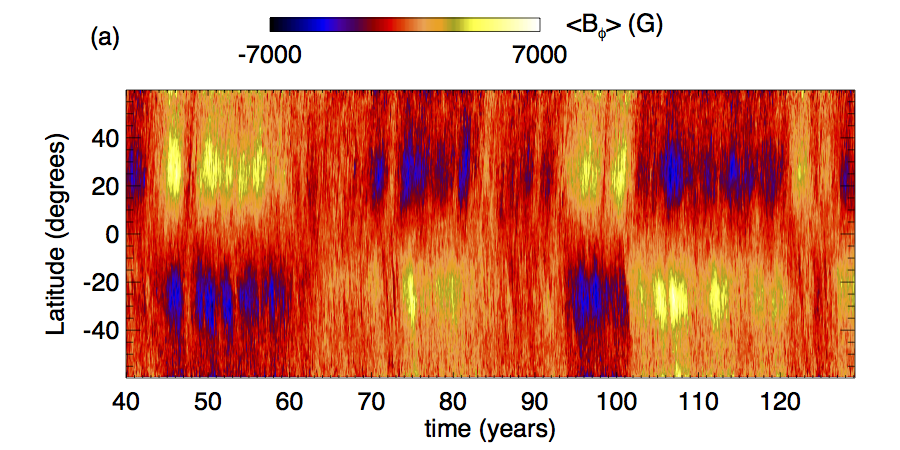} \\
\includegraphics[width=10cm,angle=0]{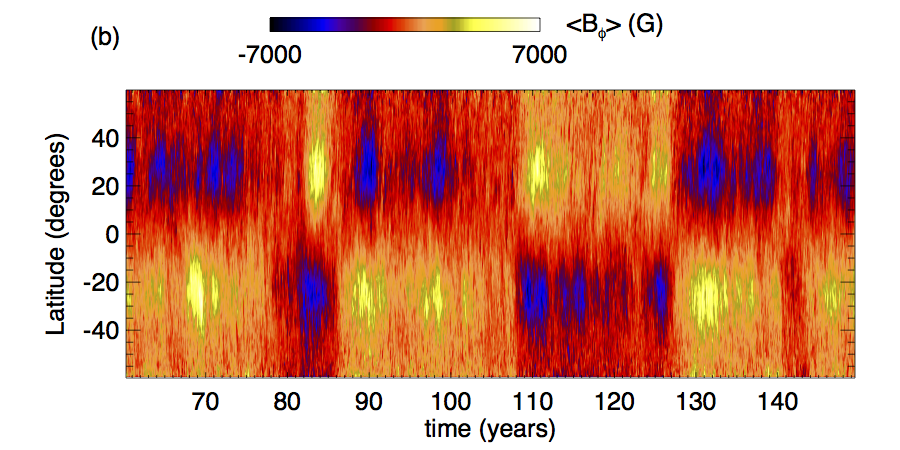}
\caption{The latitude-time diagram of the (azimuthally averaged) mean toroidal magnetic field near
the bottom of the convection zone for a sample time span of 89 years from the present convective
dynamo simulation (panel a) compared to that of a same length of time span obtained from the
convective dynamo simulation in FF (panel b).}
\label{fig:lat_time_b3}
\end{figure}

\begin{figure}
\centering
\includegraphics*[width=10cm,angle=0]{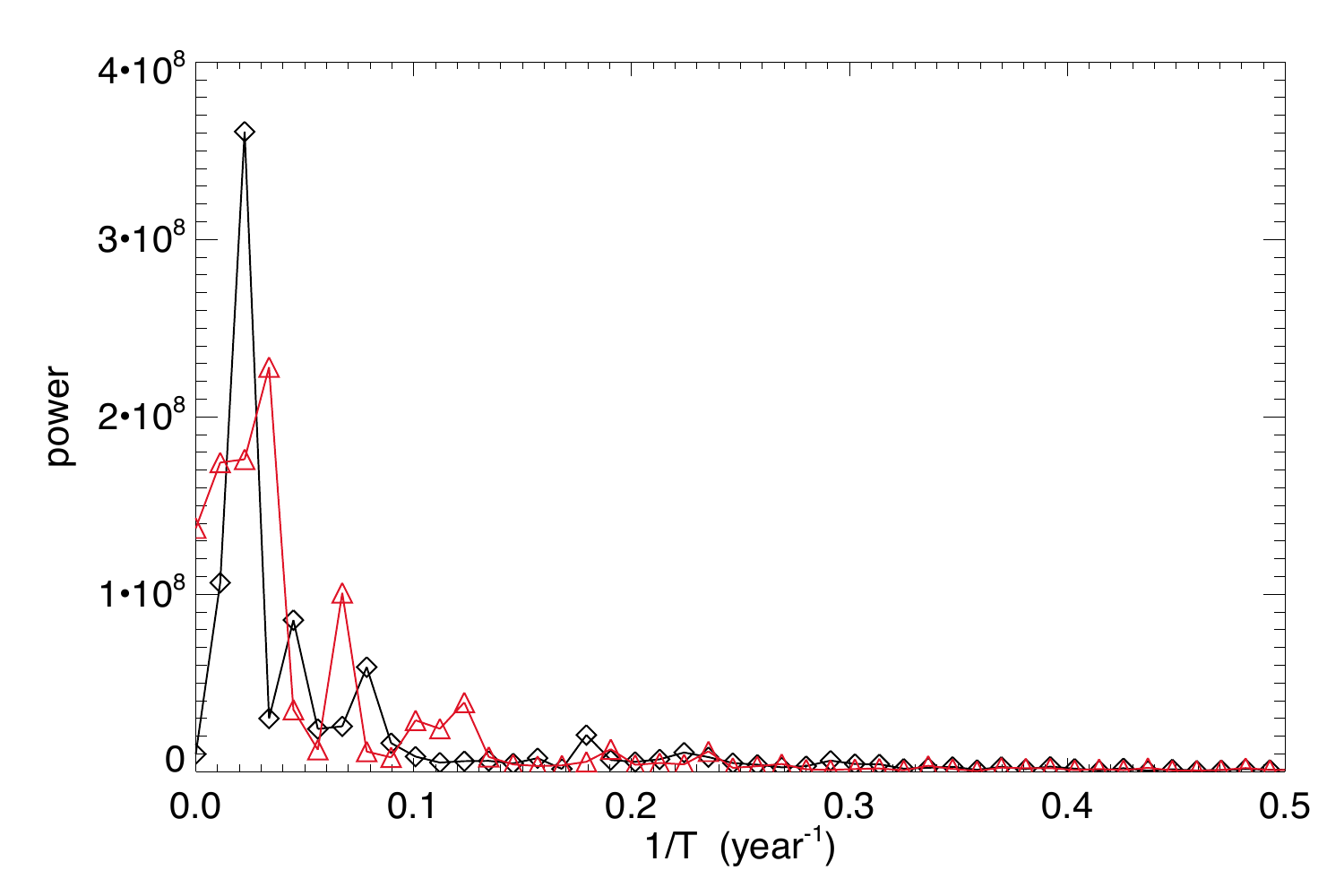}
\caption{Frequency power spectrum obtained by carrying out Fourier analysis
of the time sequence at each latitude in the latitude-time diagrams in Figure
\ref{fig:lat_time_b3}, and averaging over the latitudes. The black curve (red
curve) is for the present case (FF case).}
\label{fig:freq_b3power}
\end{figure}

The large-scale mean (azimuthally averaged) toroidal magnetic field shows
a irregular cyclic behavior
with irregular polarity reversals, as can be seen in the latitude-time diagram
of the mean toroidal magnetic field near the bottom of the convection zone
(see Figure \ref{fig:lat_time_b3}(a)), qualitatively similar to the result
in FF (shown here in Figure \ref{fig:lat_time_b3}(b)).
%%\textcolor{red}
{By carrying out Fourier analysis
of the time sequence at each latitude in the latitude-time diagrams in Figure
\ref{fig:lat_time_b3}, and averaging over the latitudes, we obtain the
frequency power spectrum as shown in Figure \ref{fig:freq_b3power}
for the present case (black curve) and for FF (red curve).
We see power peaks at periods of about 44.6 years, 22.3 years, 12.8 years,
5.6 years for the first 4 peaks in the present case.  The second dominant
peak is at the solar magnetic cycle period (about 22 years).  
For the FF case, we see power peaks at periods of about 29.8 years, 14.9 years,
8.1 years, and 5.3 years for the first 4 peaks, with the solar magnetic cycle
period falls between the 2 most dominant peaks.}
%%\textcolor{red}
{The increase in the periods for the power peaks may be related to an increase
in the magnetic diffusion time scales due to the reduction of the
magnetic diffusivity in the present case.}

\begin{figure}
\centering
\includegraphics[width=0.30\linewidth]{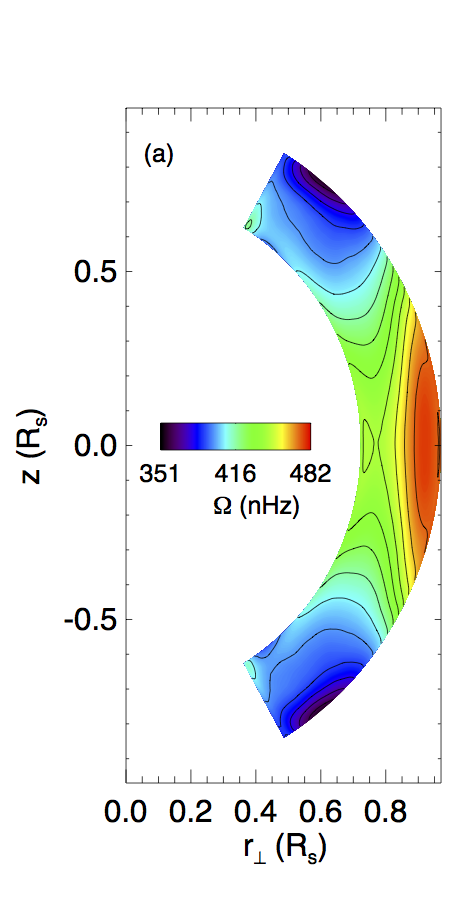}
\includegraphics[width=0.30\linewidth]{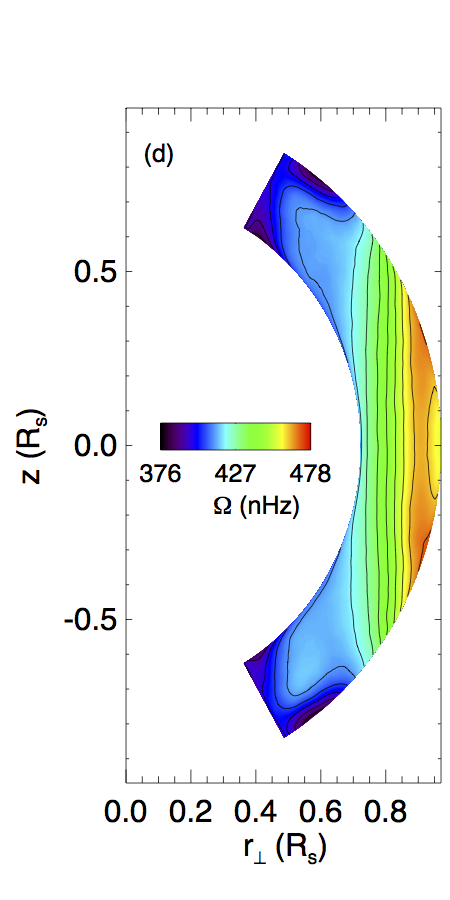} \\
\includegraphics[width=0.30\linewidth]{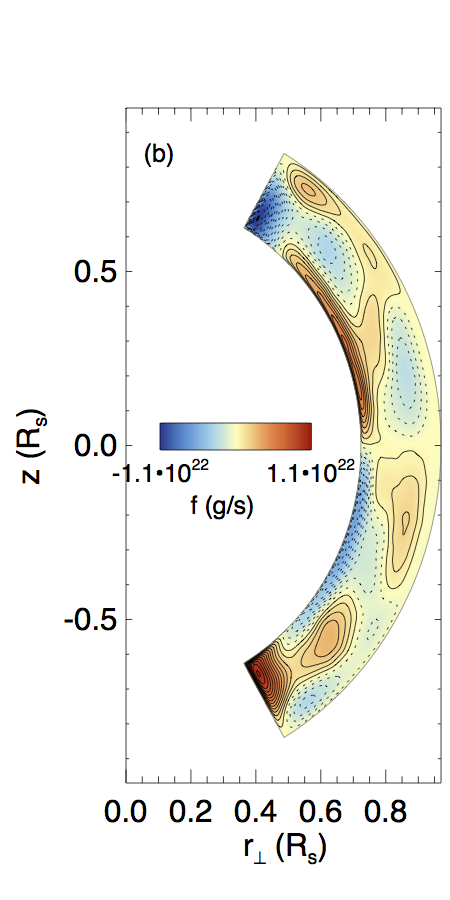}
\includegraphics[width=0.30\linewidth]{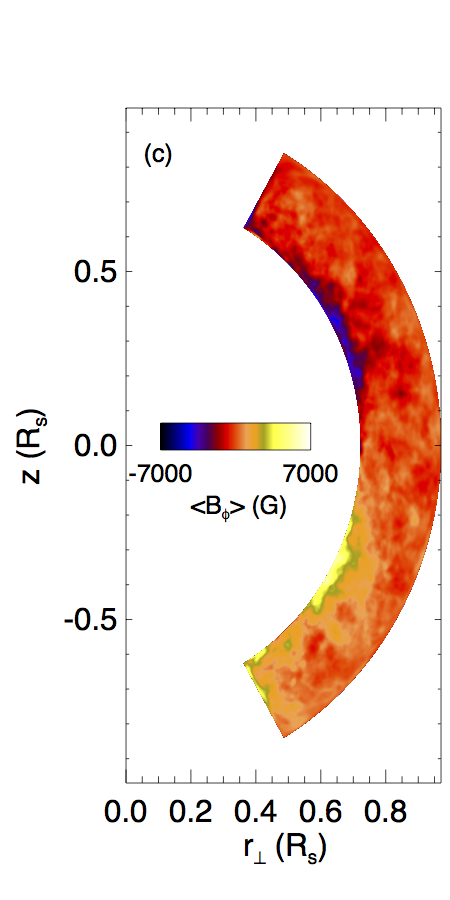}
\caption{(a) The time and azimuthally averaged angular rotation rate, (b) time
and azimuthally averaged meridional circulation mass flux $f$,
where $\rho_0 \left < {\bf v}_{m} \right > = \nabla
\times [ (f/r \sin \theta ) {\hat {\bf \phi}} ]$, and $\left < {\bf v}_m \right >$
denotes the mean velocity in the meridional plane,
(c) a snap shot of the azimuthally averaged toroidal magnetic field in the
meridional plane, and (d) the time and azimuthally averaged agular rotation
rate obtained if the latitudinal entropy gradient imposed at the bottom
boundary of the convection zone is removed.}
\label{fig:meriplanes}
\end{figure}

Figure \ref{fig:meriplanes} shows the time and azimuthally averaged angular
rotation rate (panel (a)), the time and azimuthally averaged meridional
circulation mass flux (panel (b)), and a snapshot of the azimuthally averaged
toroidal magnetic field in the meridional plane (panel (c)) resulting
from the present simualation.
The differential rotation profile self-consistently maintained by the
convective dynamo is solar-like \citep[e.g.][]{Thompson:etal:2003}, with a
faster rotation rate at the outer equatorial region than that at the polar
region by about 30\% of the mean rotation rate,
%%\textcolor{red}
{and with the isorotation
contours bending towards conical shaped in the mid-latitude zone.}
In the high and mid latitude region and at deeper depths, the iso-rotation
contours bend toward horizontal to show a radial gradient of differential
rotation, similar to the observed solar differential rotation profile.
%%\textcolor{red}
{Figure \ref{fig:meriplanes}(d) shows the resulting profile of differential
rotation if the imposed latitude gradient of entropy at the base of the
of the convection zone is removed. We see that the iso-rotational contours
become more cylindrical, and the tendency for them to turn towards conical
shaped in the mid latitude region is significantly diminished.
This confirms the role of the
tachocline induced entropy variation at the bottom of the convection
zone that can break the Taylor-Proudman constraint in the
convection zone as shown by \citet{Rempel:2005} and
\citet{Miesch:etal:2006}}.
The meridional circulation
(Figure \ref{fig:meriplanes}(b)) in the present simulation shows a
multi-cell pattern, with a main
counter-clockwise (clock-wise) circulation cell in the low latitude zone
of the northern (southern) hemisphere, producing a poleward near-surface
meridional flow at the lower latitude region.  The large-scale mean toroidal
magnetic field (Figure \ref{fig:meriplanes}(c)) is concentrated towards the bottom
of the convection zone, and is anti-symmetric in the two hemispheres.
The above results on the differential rotation, the meridional circulation,
and the large-scale mean toroidal magnetic field pattern are essentially the
same as those obtained in the simulation of FF, even though the
viscosity and magnetic diffusivity are significantly reduced.

To understand the maintenance of the solar-like differential rotation, we
examine the steady-state azimuthally averaged $\phi$-component of the 
momentum equation
%%\textcolor{red}
{(see e.g. \citet{Brun:etal:2004}, \citet{Nelson:etal:2013})},
which can be written as:
\begin{equation}
\nabla \cdot \left ( {\bf F}_{\rm RS} + {\bf F}_{\rm MC} + {\bf F}_{\rm MS}
+{\bf F}_{\rm VS} \right ) = 0
\label{eq:angularmomtrans}
\end{equation}
where ${\bf F}_{\rm RS}$, ${\bf F}_{\rm MC}$, ${\bf F}_{\rm MS}$, and
${\bf F}_{\rm VS}$ denote the angular momentum flux density due to,
respectively, the Reynolds stress (RS), the meridional circulation (MC),
the magnetic stress (MS), and the viscous stress (VS), and they are given
by the following:
\begin{equation}
{\bf F}_{\rm RS} = \rho_0 r_{\perp} \left ( \left < v'_r v'_{\phi} \right >
{\hat {\bf r}} + \left < v'_{\theta} v'_{\phi} \right >
{\hat {\bf \theta}} \right ) ,
\label{eq:f_rs}
\end{equation}
\begin{equation}
{\bf F}_{\rm MC} = \rho_0 L \left ( \left < v_r \right > {\hat {\bf r}}
+ \left < v_{\theta} \right > {\hat {\bf \theta}} \right ) ,
\label{eq:f_mc}
\end{equation}
where
%%\textcolor{red}
{$r_{\perp} = r \sin \theta$},
$L = {r_{\perp}}^2 \Omega = r_{\perp} \left ( r_{\perp} \Omega_0 + \left < v_{\phi} \right >
\right )$ is the specific angular momentum,
\begin{equation}
{\bf F}_{\rm MS} = - \frac{1}{4 \pi} r_{\perp} \left (
\left < B_r B_{\phi} \right > {\hat {\bf r}}
+ \left < B_{\theta} B_{\phi} \right > {\hat {\bf \theta}}
\right )
\label{eq:f_ms}
\end{equation}
\begin{equation}
{\bf F}_{\rm VS} = - \rho_0 r_{\perp} \nu \left (
\left < S_{r \phi} \right > {\hat {\bf r}}
+ \left < S_{\theta \phi} \right > {\hat {\bf \theta}}
\right )
\label{eq:f_vs}
\end{equation}
where
\begin{equation}
\left < S_{r \phi} \right > = r \frac{\partial}{\partial r}
\left ( \frac{ \left < v_{\phi} \right > }{r} \right ) ,
\end{equation}
\begin{equation}
\left < S_{\theta \phi} \right > = \frac{ \sin \theta}{r}
\frac{\partial}{\partial \theta} \left (
\frac{ \left < v_{\phi} \right > }{ \sin \theta } \right ) .
\end{equation}
In the above, ``$\left < \right >$'' denotes azimuthal and time averages,
superscript `` ' '' denotes the azimuthally varying component relative to the
azimuthal average,
and $r_{\perp}$ is the distance from the rotation axis.
It is insightful to examine the net angular momentum flux produced by each
of ${\bf F}_{\rm RS}$, ${\bf F}_{\rm MC}$, ${\bf F}_{\rm MS}$, and
${\bf F}_{\rm VS}$ integrated over individual concentric
cylinders $\Sigma (r_{\perp})$ of radius $r_{\perp}$.
It is approximately true that the specific angular momentum
$L \approx L(r_{\perp})$ is nearly a function of $r_{\perp}$ only, then
it can be shown that the meridional circulation produces nearly no net
angular momentum flux across each cylinder \citep[e.g.][]{Miesch:2005}, i.e.:
\begin{equation}
\int_{\Sigma (r_{\perp})} {\bf F}_{\rm MC} \cdot {\hat {\bf r}}_{\perp}
d \Sigma \approx L(r_{\perp} ) \int_{\Sigma (r_{\perp})} \rho_0
\left ( \left < v_r \right > {\hat {\bf r}}
+ \left < v_{\theta} \right > {\hat {\bf \theta}} \right )
\cdot {\hat {\bf r}}_{\perp} d \Sigma
= 0 ,
\end{equation}
because the net mass flux through each cylinder (the second integral in
the above equation) must be zero.
Thus, if there is a net angular momentum flux across the cylinders by the
Reynolds stress, then it cannot be balanced by the meridional circulation
transport and has to be balanced by the magnetic and viscous stresses, which
would generally result in the generation of differential rotation. It is
found that a net {\it outward} transport of angular momentum across the
cylinders by the Reynolds stress is an effective way of speeding up the
rotation at the equator \citep[e.g.][]{Rempel:2005}.

\begin{figure}
\centering
\includegraphics[width=1.0\linewidth]{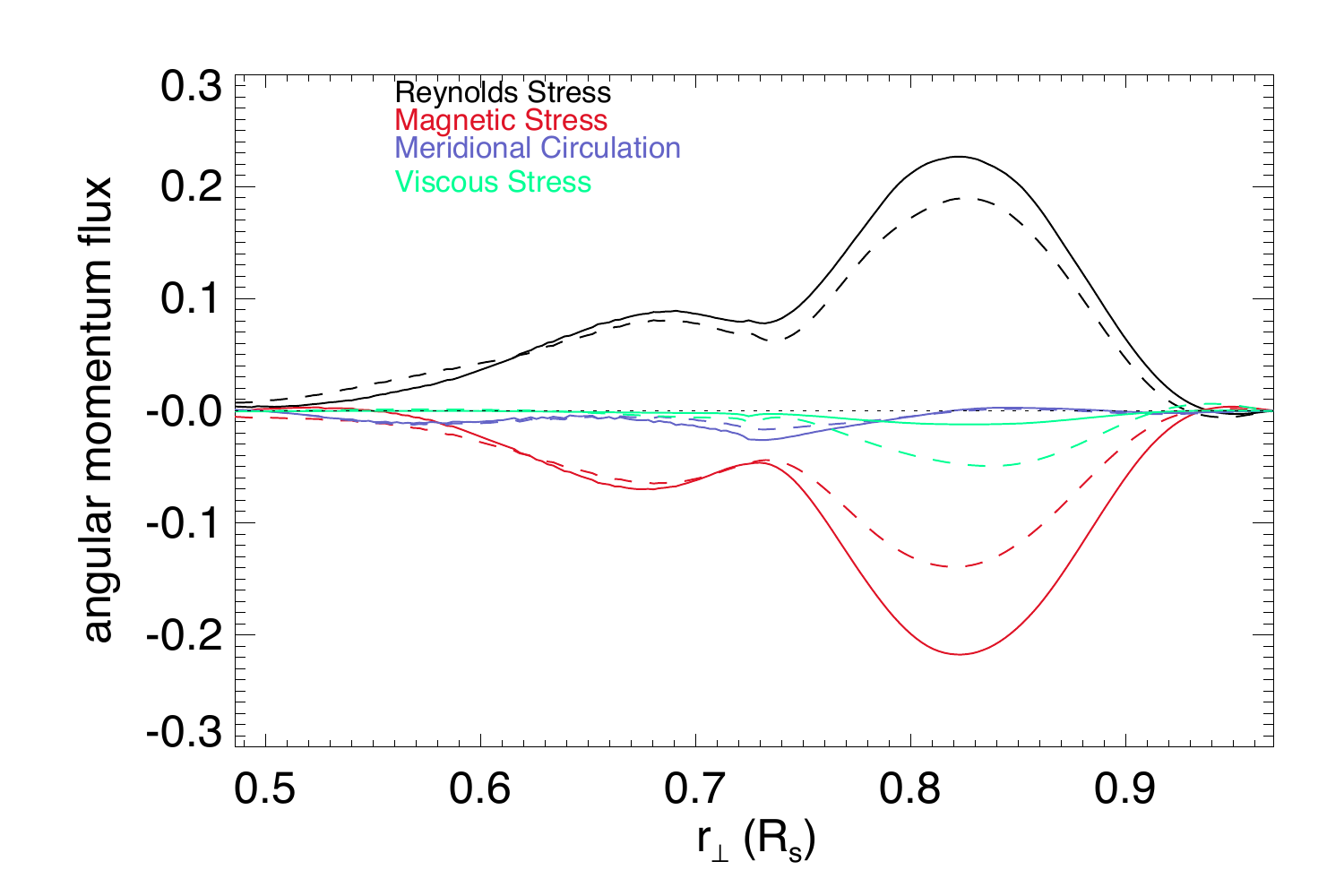}
\caption{Net angular momentum flux integrated over individual concentric
cylinders of radius $r_{\perp}$ centered on the
rotation axis produced by the Reynolds stress (black curves),
the magnetic stress (red curves),
the meridional circulation (blue curves),
and the viscous stress (green curves).
The solid curves are the results from the current simulation with reduced
viscosity and magnetic diffusivity, and the dashed curves give the
corresponding results from FF for comparison. Note here for the
computation of the viscous flux (the green curves) we have include
an estimate of the contribution from the numerical
diffusion in addition to that from the explicit viscosity given in equation
(\ref{eq:f_vs}). In the case of the solid green curve, it contains the sole
contribution from the numerical diffusion.}
\label{fig:rnrpfluxcompare}
\end{figure}

Figure \ref{fig:rnrpfluxcompare} shows the net angular momentum flux
across individual concentric
cylinders $\Sigma (r_{\perp})$ of radius $r_{\perp}$ centered on the
rotation axis produced by the Reynolds stress (black curves):
$\int_{\Sigma (r_{\perp})}
{\bf F}_{\rm RS} \cdot {\hat {\bf r}}_{\perp} d \Sigma $,
the magnetic stress (red curves):
$\int_{\Sigma (r_{\perp})}
{\bf F}_{\rm MS} \cdot {\hat {\bf r}}_{\perp} d \Sigma $,
the meridional circulation (blue curves):
$\int_{\Sigma (r_{\perp})}
{\bf F}_{\rm MC} \cdot {\hat {\bf r}}_{\perp} d \Sigma $,
and the viscous stress (green curves):
$\int_{\Sigma (r_{\perp})}
{\bf F}_{\rm VS} \cdot {\hat {\bf r}}_{\perp} d \Sigma $.
The solid curves are the results from the current simulation with reduced
viscosity and magnetic diffusivity, and the dashed curves give the
corresponding results from FF for comparison.
%%\textcolor{red}
{Note here for the green curves showing the viscous flux,
we have included in the computation of 
${\bf F}_{\rm VS}$ an estimate of the contribution due to the numerical
diffusion in addition to that due to the explicit viscosity given by equation
(\ref{eq:f_vs}). In the case for the solid green curve in Figure \ref{fig:rnrpfluxcompare}, it contains the sole contribution from the numerical diffusion.}
From Figure \ref{fig:rnrpfluxcompare} we see that indeed the meridional
circulation contributes very little to
the net transport of angular momentum across the cylinders.  We see a
net outward (positive) transport of angular momentum by the Reynolds stress
(black curves) throughout the
convection zone (except at the very top), which is the driver and cause of
the solar-like (faster equator) differential rotation in the convective dynamo.
%%\textcolor{red}
{Similar results are also found in \citet{Nelson:etal:2013}}.
In FF we found that the presence of the magnetic fields in the convective
dynamo is important for the self-consistent maintenance of the
solar-like differential rotation. Without the magnetic
fields, in the corresponding hydrodynamic case, the net transport of angular
momentum by the Reynolds stress reverses to inward across the cylinders,
resulting in an anti-solar (faster pole) rotation profile.
%%\textcolor{red}
{(This result of changing to an anti-solar rotation with the removal of
the magnetic field does not depend on whether the latitude
gradient of entropy is imposed at the bottom of the convection zone.)}
To get a solar-like differential rotation in the hydrodynamic case, a much
higher viscosity (by about 5 times the viscosity used in the convective dynamo
case in FF) is needed to suppress the convection and obtain an outward
transport of angular momentum by the Reynolds stress.
In the present convective dynamo simulation where we have reduced the viscosity
and the magnetic diffusivity compared to FF, we find that the transport
of angular momentum by the Reynolds stress remains outward 
and its amplitude is enhanced for most of the convection zone compared to
the convective dynamo in FF (compare the black solid and black dashed curves
in Figure \ref{fig:rnrpfluxcompare}).
In turn, the inward angular momentum flux by the magnetic stress is also
increased (compare the solid and the dashed red curves in Figure
\ref{fig:rnrpfluxcompare}) to nearly balance the increased Reynolds stress
flux, and the net angular momentum flux by the viscous stress is significantly
reduced to a negligible level (compare the solid and dashed green curves) because
of the removal of the explicit viscosity.
The small amount of the remaining flux due to viscous stress as shown by the
solid green curve is computed from the numerical diffusion.
Here again we see the dynamical effect of the magnetic fields which takes
up an effective role of a viscosity.

\begin{figure}
\centering
\includegraphics[width=1.0\linewidth]{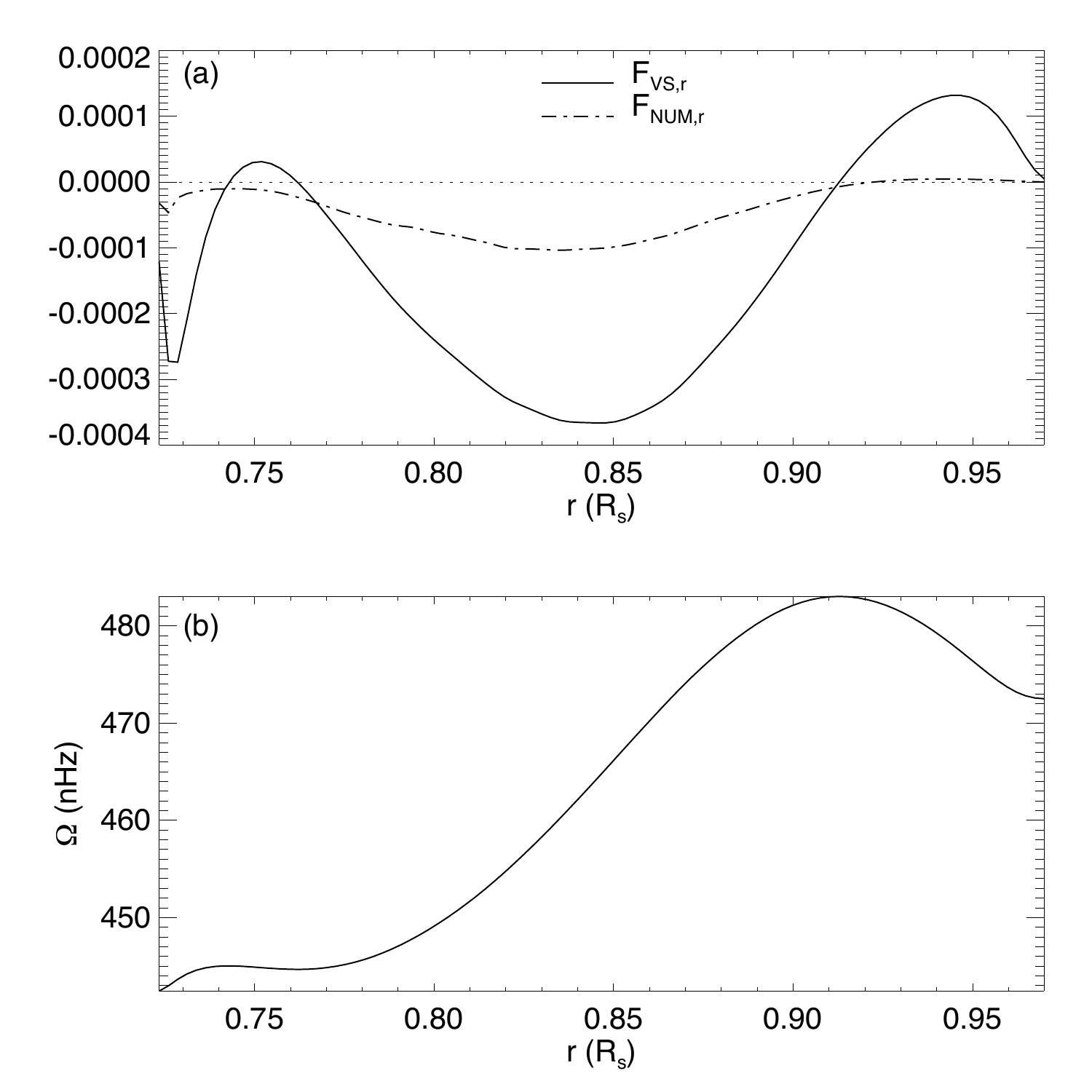}
\caption{(a) The radial angular momentum flux density $F_{\rm VS, r}$ at
the equator due to the explicit viscosity (solid curve),
compared to the corresponding (time and azimuthally averaged) radial angular
momentum flux density $F_{\rm NUM, r}$ due to numerical diffusion
(dash-dotted curve),
%%\textcolor{red}
{both computed in the simulation of FF}.
(b) The time and azimuthally averaged angular
frequency $\Omega$ of the differential rotation at the equator (also from the FF case).}
\label{fig:vsrp_numrp_equator}
\end{figure}

%%\textcolor{red}
{To assess the relative importance of the numerical and explicit viscosities
on smooth, well resolved flows in the case of FF, we have computed in Figure
\ref{fig:vsrp_numrp_equator}a the radial angular momentum flux density due to
the explicit viscosity, evaluated at the equator:
\begin{equation}
F_{\rm VS, r}=-\rho_0 r \nu \left < S_{r \phi} \right >
=- \rho_0 r^2 \nu \frac{\partial \Omega}{\partial r}
\label{eq:F_VS_r}
\end{equation}
and the corresponding (time and azimuthally averaged) radial
angular momentum flux density $F_{\rm NUM, r}$ due to
numerical diffusion produced by the upwinded evaluation of the
advection term.  We see that the amplitude of $F_{\rm NUM, r}$ is typically
smaller than that of $F_{\rm VS, r}$ by more than a factor of 3.  This suggests
that the effective numerical viscosity acting on the well-resolved, smooth
variation of $\Omega$ (Figure \ref{fig:vsrp_numrp_equator}b) is significantly
smaller (by typically a factor of 3) than the explicit viscosity $\nu$
%%\textcolor{red}
{in the FF case}.
The numerical diffusion changes with the local spatial properties of the
flows. For non-monotonic grid-scale fluctuations, the numerical diffusivity
dominates.  But for well resolved flows such as the smoothly varying $\Omega$,
we find that the numerical diffusivity is significantly smaller than the
explicit viscosity, and thus removing the explicit viscosity in the
current case is making a significant difference for the diffusion on the
well resolved scales.}

\begin{figure}
\centering
\includegraphics[width=0.75\linewidth]{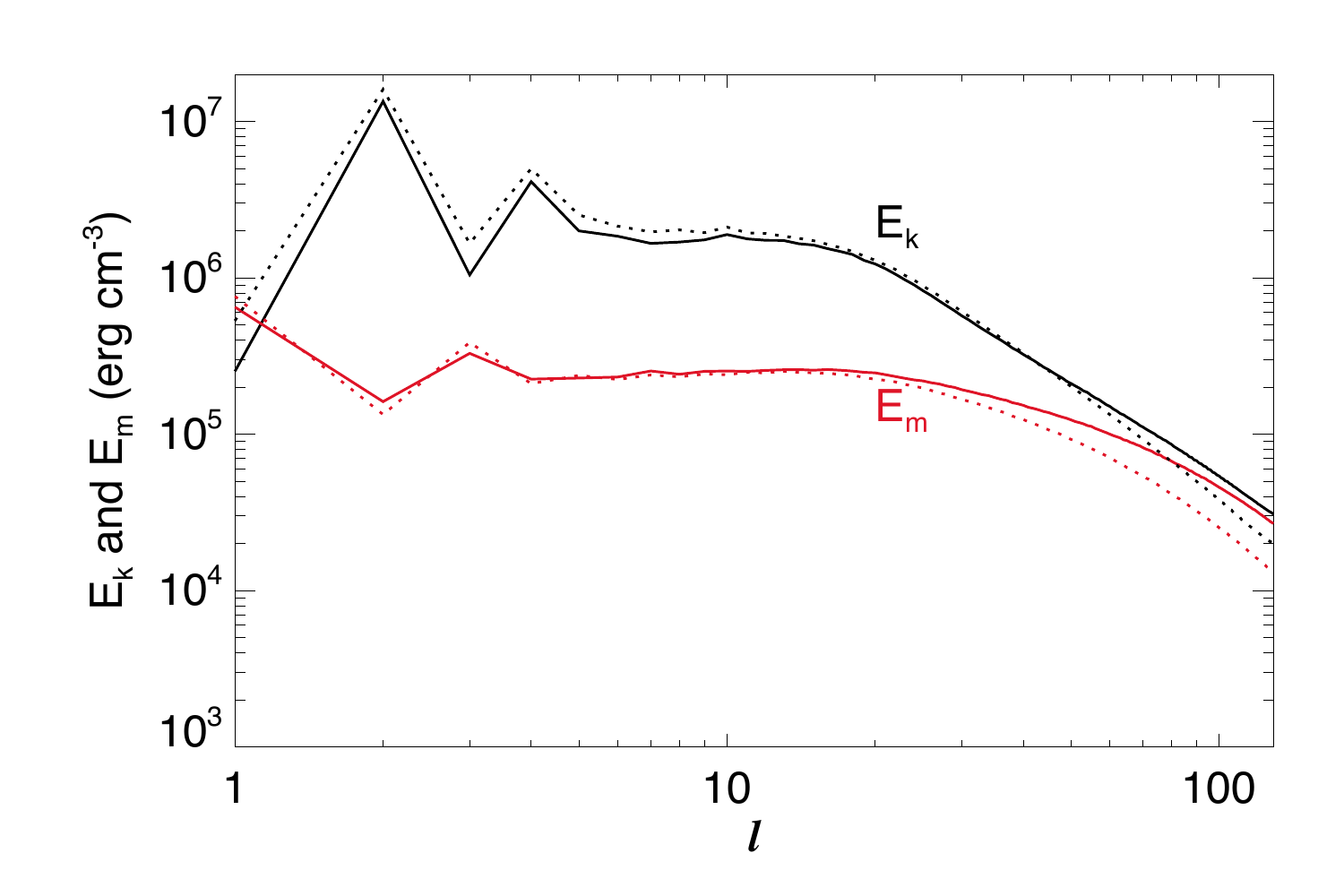} \\
\includegraphics[width=0.75\linewidth]{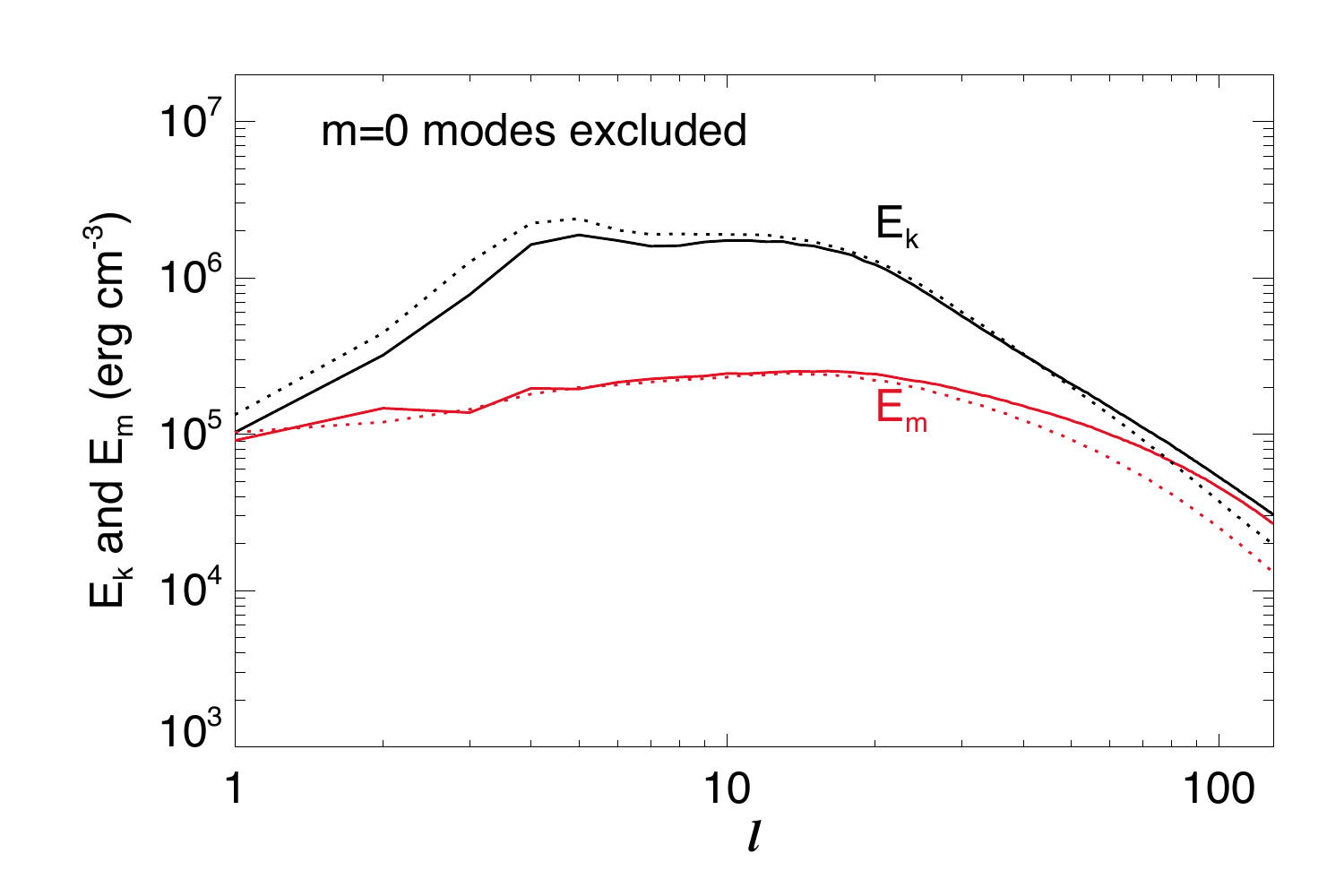}
\caption{The top panel shows the kinetic ($E_k$, black curves) and
magnetic ($E_m$, red curves)
energy spectra as a function of the spherical harmonic degree $l$ at a depth
of $0.73 R_{\odot}$ near the bottom of the convection zone from our
current convective dynamo simulation (solid curves) with reduced viscosity
and magnetic diffusivity, and from the convective dynamo of FF (dotted lines).
The bottom panel shows the same but with the contribution from the mode with
the azimuthal order $m=0$ taken out for each $l$.}
\label{fig:energyspectra}
\end{figure}

The reason for the enhanced outward angular momentum flux by the Reynolds
stress (as seen in Fig. \ref{fig:rnrpfluxcompare}) in the current case with
reduced viscosity and magnetic diffusivity is
due to the enhanced magnetic fields, which further suppress
convection and makes it more rotationally constrained.
The top panel of Figure \ref{fig:energyspectra} shows the
kinetic (black curves) and magnetic (red curves) energy spectra
as a function of the spherical harmonic degree $l$ at a depth of
$0.73 R_{\odot}$ near the bottom of the convection zone from the present
convective dynamo (solid curves) and from the 
convective dynamo of FF (dotted curves).
We show energy spectra up to the highest $l$ (about 128) that is
adequately resolved.
%%\textcolor{red}
{The bottom panel of Figure \ref{fig:energyspectra} shows the same as the
top panel but with the contribution from the mode with azimuthal order $m=0$
taken out for each $l$.  As a result the large contribution from the
differential rotation, which produces the two peaks in the kinetic energy
seen at $l=2$ and $l=4$ in the top panel, are removed such that the kinetic
energy spectra (in the bottom panel) reflects more closely those for the
convective motions.
Also the contribution from the (azimuthally averaged) mean magnetic field
is removed from the magnetic energy spectra in the bottom panel.
Compared to FF, we see that the magnetic energy $E_m$ (in both the top and
bottom panels) in the
present case is enhanced significantly in the small
scales and remains about the same in the large scales.
The convective kinetic energy on the other hand is reduced in the larger scales,
which are the ones more influenced by rotation and contributing to the
Reynolds stress transport of angular momentum.
These larger scale convective motions become more
rotationally constrained due to the reduction in their amplitude and thus
producing a greater
outward flux of angular momentum by the Reynolds stress.
In the smaller scales, the kinetic energy is increased compared to FF, but
the magnetic energy is also increased and becomes closer to equipartition with
the kinetic energy compared to FF. Thus the magnetic field becomes more
dynamically important in the present case than in FF.}
%%We also computed the Rossby number defined in the same way as that in FF
%%and found that it is $0.71$ compared to $0.74$ in FF due to a slight
%%decrease of the convective RMS velocity in the present case.  

We have computed the following energy conversion rates, the
buoyancy work $W_{\rm Buoyancy}$, which is the energy source and driving
of the convection, and the work done against the Lorentz
force $W_{\rm Lorentz}$, which measures
the energy conversion from the kinetic energy to the magnetic energy:
\begin{equation}
W_{\rm Buoyancy} = \left < \int_V \rho_0 g_0 v_r \frac{s_1}{c_p} dV \right > ,
\label{eq:buoyancywork}
\end{equation}
\begin{equation}
W_{\rm Lorentz} = - \left < \int_V {\bf v} \cdot \left [ \frac{1}{4 \pi}
\left ( \nabla \times {\bf B} \right ) \times {\bf B} \right ] dV \right >,
\label{eq:work_against_lorentz}
\end{equation}
where the integration is over the entire simulation volume $V$, and
``$\left < \right >$'' denotes time average.  
The buoyancy work $W_{\rm Buoyancy}$ is found to be nearly identical for
the two cases: being $0.782 \pm 0.0023 L_{\rm sol}$
for the present simulation and $0.777 \pm 0.0024 L_{\rm sol}$ for FF,
where $L_{\rm sol}$
is the solar luminosity, with the difference being below the 3 sigma level.
In other words, the convection is driven (nearly) equally hard in the two
convective dynamo simulations.  This is probably because both are driven by
the same solar radiative diffusive heat flux and we have used the same
thermal diffusivity. On the other hand, the work done against the Lorentz
force $W_{\rm Lorentz}$ is $0.330 \pm 0.0074 L_{\rm sol}$ for the present
case and $0.256 \pm 0.0067 L_{\rm sol}$ for FF, i.e. with more kinetic energy
converted to the magnetic energy in the present case.
Thus, as a result of the reduced viscosity and magnetic diffusion in our
present simulation, an
increased portion of the (same) buoyancy work is being converted to the
magnetic energy, and a decreased portion for work against viscous dissipation.
This is another way to see the effective role of the magnetic fields as an
enhanced viscosity in suppressing convection.

\begin{figure}
\centering
\includegraphics[width=1.0\linewidth]{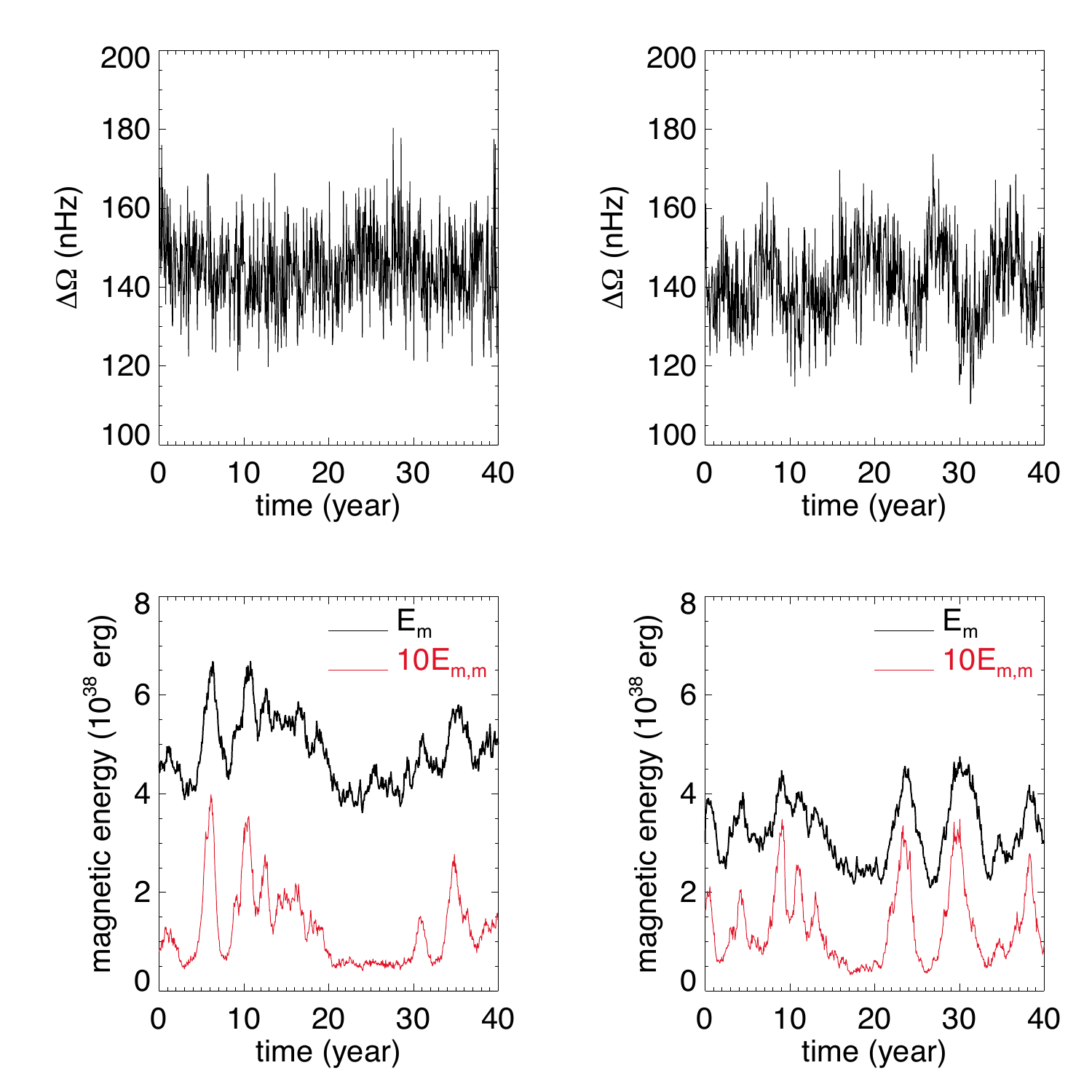}
\caption{Top panels show the evolution of the differential rotation
contrast $\Delta \Omega$ (peak difference in angular rotation rate)
and the bottom panels show the corresponding temporal evolution of the total
magnetic energy $E_{\rm m}$ (black curve) and ten times the mean magnetic energy
$10 E_{\rm m,m}$ (red curve),
for the present low diffusivity case (left column) and for
FF (right column).}
\label{fig:deltaomg_evol}
\end{figure}

%%\textcolor{red}
{Figure \ref{fig:deltaomg_evol} shows the temporal evolution of the
differential rotation contrast $\Delta \Omega$ (peak difference in angular rotation rate, top panels) and
%%\textcolor{red}
{the corresponding temporal evolution of the total and (10 times) the
mean magnetic energies (bottom panels)},
for the present convective dynamo (left column) in comparison to the case
of FF (right column).  $\Delta \Omega$ on average is slightly bigger, being
$144.1 \pm 0.2 $ nHz in the present case, compared to $142.5\pm0.3$ nHz for FF.
$\Delta \Omega$ shows a temporal variation with an RMS amplitude of about
$10$ nHz for both cases, weakly anti-correlated with the total magnetic energy
$E_{\rm m}$
%%\textcolor{red}
{as well as the mean magnetic energy $E_{\rm m,m}$}.
The anti-correlation is weaker in the present case (with a linear correlation
coefficient $r=-0.21$
%%\textcolor{red}
{with $E_m$ and $r=-0.15$ with $E_{\rm m,m}$}
) compared to FF ($r=-0.44$
%%\textcolor{red}
{with $E_m$ and $r=-0.33$ with $E_{\rm m,m}$}
). The magnetic field is
playing a complex dual role for the differential rotation.  On the one hand it
suppresses convection to allow a greater outward transport of the angular
momentum by the Reynolds stress to drive a solar-like differential
rotation.  On the other hand it also damps the differential rotation by
balancing the Reynolds stress with the
Maxwell stress transport of angular momentum. Such complex dual role may
have resulted in a weak anti-correlation.}

\begin{figure}
\centering
\includegraphics[width=10cm]{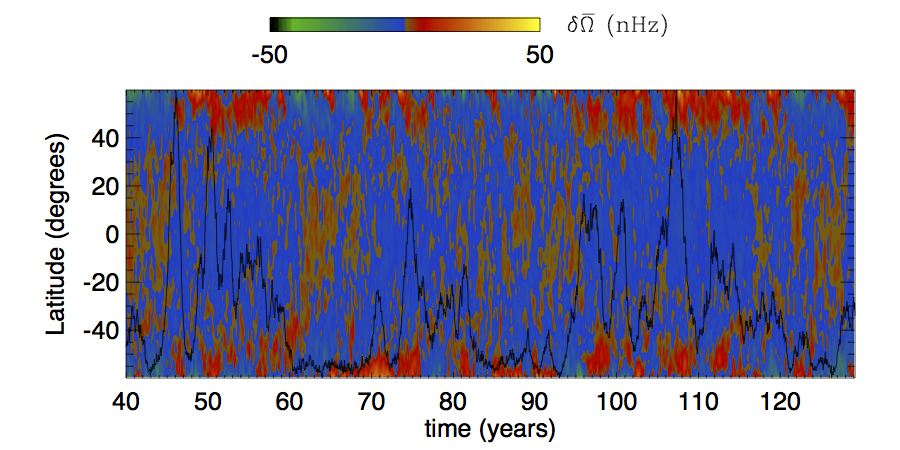} \\
\includegraphics[width=10cm]{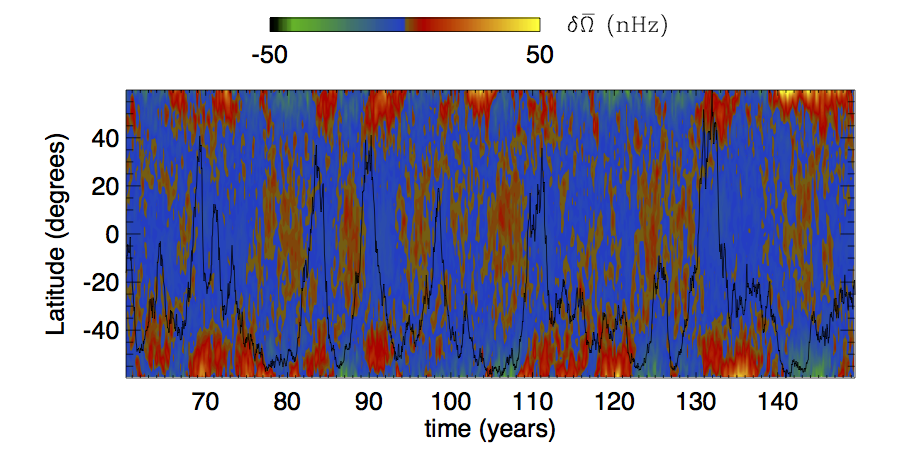}
\caption{%%\textcolor{red}
{The latitude-time diagram of
$\delta \bar{\Omega} (t, \theta )= \bar{\Omega} (t, \theta ) -
\left < \bar{\Omega} \right >_t (\theta )$, where $\bar{\Omega} (t, \theta)$
is the depth averaged angular rotation rate as a function of time and latitude,
and $\left < \bar{\Omega} \right >_t (\theta )$ is the temporal
average of $\bar{\Omega} (t, \theta)$.
Thus $\delta \bar{\Omega} (t, \theta )$ shows
the deviation from the mean latitudinal differential rotation. The overlaid
black curve shows the temporal evolution of the mean magnetic energy.
The top panel shows the results for the present low diffusivity case and the
bottom panel shows the results from the FF simulation.}}
\label{fig:lat_time_omg}
\end{figure}

%%\textcolor{red}
{Figure \ref{fig:lat_time_omg} shows the latitude-time diagram of
the deviation from the (time averaged) mean latitudinal differential rotation,
i.e.  $\delta \bar{\Omega} (t, \theta )= \bar{\Omega} (t, \theta ) -
\left < \bar{\Omega} \right >_t (\theta )$, where $\bar{\Omega} (t, \theta)$
denotes the depth averaged angular rotation rate as a function of time and latitude,
and $\left < \bar{\Omega} \right >_t (\theta )$ is the temporal
average of $\bar{\Omega} (t, \theta)$.
Overlaid on the diagram is the (black) curve
showing the temporal variation of the mean magnetic energy $E_{\rm m,m}$.
The top panel shows
the results for the present low diffusivity case and the bottom panel shows
the results from the FF simulation.
The main pattern we see is that there is a tendency for
negative $\delta \bar{\Omega}$ at the equator and
positive $\delta \bar{\Omega}$ in the polar region for magnetic maxima, i.e.
the contrast of the (solar-like) latitudinal differential rotation is relatively
reduced for magnetic maxima.
This anti-correlation is quite weak: we find a linear correlation
coefficient between $\delta \bar{\Omega}$ at the equator and $E_{\rm m,m}$ of
$r=-0.46$ for the present case and $r=-0.48$ for the FF case.}

\begin{figure}
\centering
\includegraphics[width=1.0\linewidth]{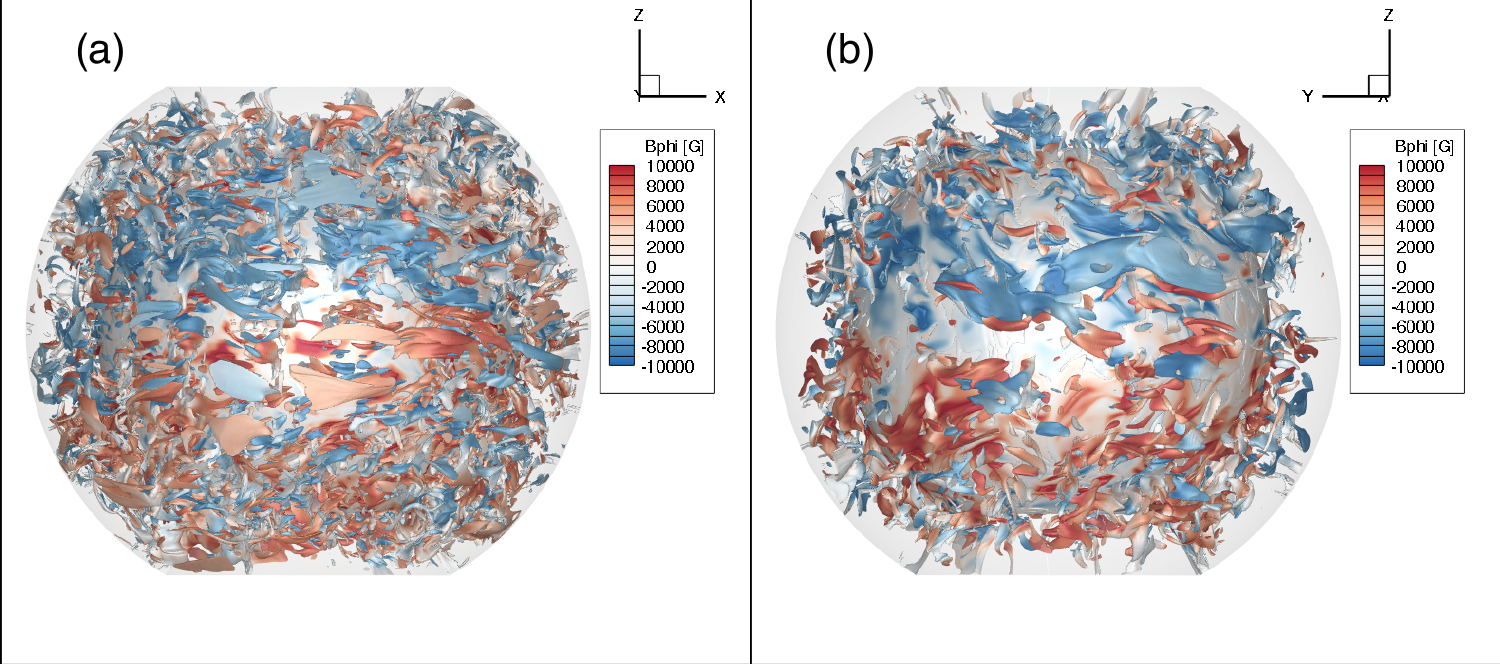}
\caption{3D views of iso-surfaces of $v_a / v_{\rm rms} = 1$ in the convective
envelope, where $v_a$ is the Alfv{\'e}n speed and $v_{\rm rms}$ is the RMS
convective velocity at that depth. The iso-surfaces outline regions of strong magnetic
field concentrations of super-equipartion field strength. (a) results from
the current less diffusive convective dynamo simulation and (b) results from the
convective dynamo simulation of FF.}
\label{fig:aovv}
\end{figure}

\begin{figure}
\centering
\includegraphics[width=1.0\linewidth]{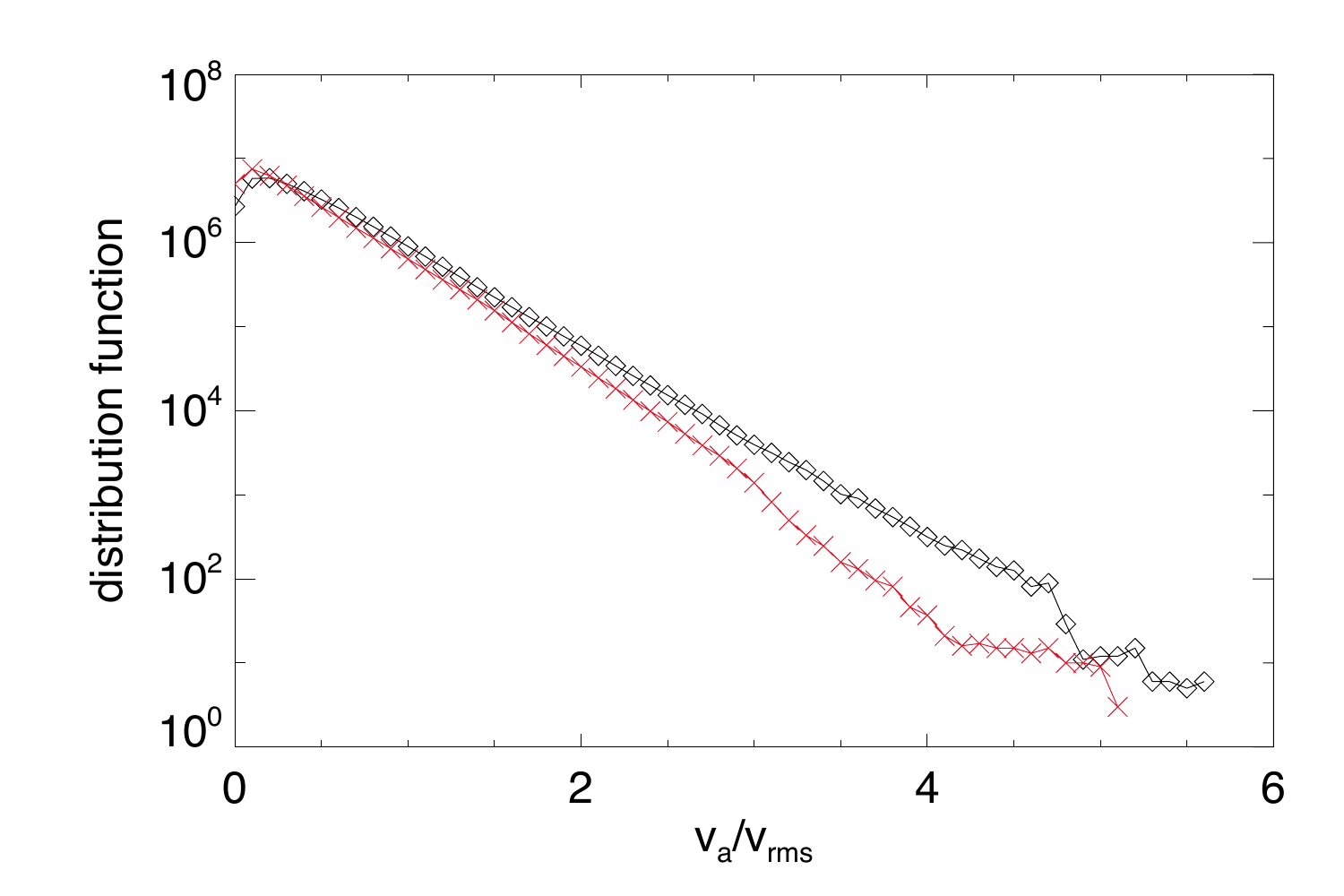}
\caption{Distribution function of the grid points over the value of $v_a / v_{\rm rms}$, where $v_a$ is the Alfv{\'e}n speed and $v_{\rm rms}$ is the RMS
convective velocity at that depth. Diamond points are the result for the
current less diffusive convective dynamo and the red cross points are the
result from FF}
\label{fig:aovv_pdf}
\end{figure}

Figure \ref{fig:aovv} shows a 3D view of the magnetic field concentrations in
the convective envelope produced by the present convective dynamo simulation
(panel (a)), in comparison to that produced by the convective dynamo of FF
(panel (b)).  The images show iso-surfaces of $v_a / v_{\rm rms} = 1$ where
$v_a$ is the Alfv{\'e}n speed and $v_{\rm rms}$ is the RMS velocity at
that depth. The iso-surfaces thus outline regions of strong magnetic field
concentrations with super-equipartition field strength. The iso-surfaces are
colored with the azimuthal field strength $B_{\phi}$ as indicated by the
color table.
We see filamentary strong field concentrations of a preferred
sign of toroidal field in each hemisphere, antisymmetric for the two hemispheres.
Some of these strong filaments emerge towards the top boundary (appearing
flattened at the top).
Comparing panels (a) and (b) we see that the present, less diffusive
convective dynamo shows denser and more smaller scale filaments of strong
(super-equipartion) magnetic field concentrations.
%%\textcolor{red}
{Figure \ref{fig:aovv_pdf} shows the distribution function of the grid points over the value of $v_a / v_{\rm rms}$, where $v_a$ is the Alfv{\'e}n speed
and $v_{\rm rms}$ is the RMS convective velocity at that depth, at a
magnetic maximum (corresponding to the snapshot shown in Figure \ref{fig:aovv}(a))
for the current less diffusive case (black diamond points)
compared to that at a magnetic maximum (corresponding to Figure \ref{fig:aovv}(b))
for the FF case (red cross points).
There is clearly more volume occupied by strong super-equipartition fields
(with $v_a / v_{\rm rms} > 1$) in the current less diffusive case, indicating
that the magnetic field is more dynamically important.}

\section{Conclusions}
\label{sec:conclusions}
Extending upon the work of FF, we investigate further the dynamical
effect of the magnetic fields in the self-consistent maintenance of
the solar-like differential rotation by carrying out a simulation of
the solar convective dynamo as that described in FF but with further
reduced viscosity and magnetic diffusivity.
The explicit viscosity and magnetic diffusion in the
simulation of FF are removed and only the slope-limited numerical
diffusions are present in the current simulation.

%%\textcolor{red}
{As a result, we find that the magnetic energy is enhanced by about 1.5
times, and mainly in the small scales, while the kinetic energy is reduced
in the larger scales and increased in the smaller scales, with the total
kinetic energy showing a decrease.
In the smaller scales, the magnetic energy is increased to approaching
equipartition with the kinetic energy.}
Because of the further suppression of the amplitude of the
larger scale convective motions which are most influenced by
rotation, they become more rotationally constrained and we found
a further increased net outward transport of angular momentum (across
the cylinders, away from the rotation axis) by the Reynolds stress, which
is in turn balanced by a significantly increased inward transport of the
angular momentum by the magnetic stress. The net angular momentum
transport by viscous stress is significantly reduced to a negligible level,
with the transport by the Reynolds stress nearly entirely
balanced by the transport by the magnetic stress.
(The net transport by the meridional circulation is also found to be
negligible.)
The resulting solar-like differential rotation 
remains essentially the same as that found in FF.
The large scale mean field also shows a similar irregular cyclic behavior
as that found in FF.
%%\textcolor{red}
{We note that our results here and in FF are limited to the case of
magnetic Prandtl number of 1. We have also only run the dynamo simulations
with a radial magnetic field top boundary condition (which has zero torque from the
top boundary to ensure angular momentum conservation of the spherical shell).}
The result presented in this convective dynamo simulation as an extension of
that in FF further demonstrates the dynamical effect of the
magnetic field that behaves likes an enhanced viscosity to suppress the
convective motions and produce the necessary Reynolds stress transport 
for maintaining the solar-like differential rotation, as the
diffusivities are reduced.
%%Recent unpresedented high-resolution global simulations of solar convective
%%dynamo at much more reduced viscosity and magnetic diffusivity
%%\citep{Hotta:etal:2015} have
%%found an efficient small-scale dynamo that suppresses small-scale flows
%%through lorentz force feedback, and thus allow the construction of the
%%large-scale cyclic mean field seen in the large diffusivity cases.
%%Our current study shows the dynamical effect of the magnetic field
%%as an enhanced viscosity in maintaining the solar-like differential rotation.

\section*{Acknowledgement}
This work is supported in part by the NASA LWSCSW grant NNX13AG04A to NCAR. NCAR
is sponsored by the National Science Foundation. F.F is supported by the
University of Colorado George Ellery Hale Postdoctoral Fellowship. The
numerical simulations were carried out on the Pleiades supercomputer at the
NASA Advanced Supercomputing Division under project GIDs s1362 and also on
the Yellowstone supercomputer at NCAR-Wyoming Supercomputing Center
(ark:/85065/d7wd3xhc) provided by NCAR's Computational and Information Systems
Laboratory, sponsored by the National Science Foundation.

%%%%%%%%%%%%%%%%%%%%%%%%%%%%%%%%%%%%%%%%%%%%%%%%%%%%%%%%%%%%%%%%%%%%%%%%%%%%%
%% Appendices
% \appendix

%%\clearpage
%%\bibliographystyle{model5-names}\biboptions{authoryear}
%%\bibliography{references}

%%\begin{table}
%%\caption{This is the caption of this table}
%%\begin{tabular}{ll}
%%\hline
%%Parameter&Value\\
%%\hline
%%Parameter 1 & $526.849 \pm 0.003$ s \\
%%Parameter 2 & $5268.49 \pm 0.03$ s \\
%%Parameter 3 & $52684.9 \pm 0.3$ s \\
%%\hline
%%\end{tabular}
%%\label{table1}
%%\end{table}

\clearpage

\clearpage

\clearpage

\clearpage

\clearpage

\clearpage

\clearpage

\clearpage

\clearpage

\clearpage

\clearpage

\end{document}